**Quickly fading afterimages: hierarchical adaptations in human perception**

Madeline E. Klinger[1,2], Christian A. Kell[2], Danko Nikolić[3,4,5,6]

1) University of California, Berkeley, CA, USA
2) Department of Neurology and Brain Imaging Center, Goethe University Frankfurt, Germany
3) Frankfurt Institute for Advanced Studies, Frankfurt, Germany
4) Max Planck Institute for Brain Research, Frankfurt, Germany
5) Department of Psychology, University of Zagreb, Croatia
6) savedroid AG, Frankfurt, Germany

Correspondence:

Danko Nikolić
Frankfurt Institute for Advanced Sciences
Ruth-Moufang-Straße 1,
60438 Frankfurt am Main, Germany
danko.nikolic@gmail.com

**Abstract**

Afterimages result from a prolonged exposure to still visual stimuli. They are best detectable when viewed against uniform backgrounds and can persist for multiple seconds. Consequently, the dynamics of afterimages appears to be slow by their very nature. To the contrary, we report here that about 50% of an afterimage intensity can be erased rapidly—within less than a second. The prerequisite is that subjects view a rich visual content to erase the afterimage; fast erasure of afterimages does not occur if subjects view a blank screen. Moreover, we find evidence that fast removal of afterimages is a skill learned with practice as our subjects were always more effective in cleaning up afterimages in later parts of the experiment. These results can be explained by a tri-level hierarchy of adaptive mechanisms, as has been proposed by the theory of practopoiesis.

To recognize the person hidden in Figure 1A, one needs to create an afterimage first. This is done by fixating on the image for about 10 seconds and then observing its 'negative' over a white background. Such afterimages usually last for many seconds (Newsome 1971; Virsu & Laurinen 1977; Robertson & Fry 1937; Ritschel & Eisemann 2012) and even minutes (Hansel & Mahmud 1978; Jameson et al. 1979). Previous studies reported fast adaptations (Rinner & Gegenfurtner 2000) as well as the fact that attention to stimuli and relevance of stimuli can affect afterimages (Baijal & Srinivasan 2009; Keller et al. 2017). We use afterimages to isolate adaptation mechanism that respond to color filters; that way the reported intensities of afterimages serve as measures of the degree to which adaptations took place. In the present study, we investigated what is required for afterimages to fade quickly.

We induced afterimages using photographs of natural scenes (Figure 1B), which were passed through a green filter (80% filtered; a series of 12 images was shown, one second each for a total of 12 seconds). This allowed us to create afterimages in form of a pink haze, for which it was then easy to give accurate reports on their pinkness. Subjects used a pink level palette to indicate the intensity of the afterimage (Figure 1B). Importantly, they made these reports at different durations of the recovery phase, ranging from 0 milliseconds (no recovery time allowed at all) up to an 800-millisecond delay. This allowed us to construct afterimage *decay curves*.

Most important for the present study were the manipulations of the contents viewed during the recovery. One condition complied with the classical studies and had subjects simply stare at a blank white screen. However, in the other conditions, subjects were presented with visual materials during recovery. These materials were not green-filtered but shown in their full colors, presuming that the visual system will remove afterimages by readjusting to the new perceptual conditions. These recovery stimuli were also created from natural scenes and contained either a single full-colored image, a single gray-scale image or a sequence of full-color images arranged in a barrage (a series of 200-millisecond long flashes) (see Figure 1B and methods).

The study was motivated by the tri-level adaptive hierarchy theory, known as practopoiesis (Nikolić 2015). This theory proposes that the fast adaptive mechanisms of nerve cells (Chung & Nelson 2002; Pearson 2000; Verhoef et al. 2008), play a critical role in perceptual and cognitive operations. We presumed here that fast cellular adaptive mechanisms are engaged while forming afterimages and tested the hypothesis that the same fast adaptations play a role in the cleanup of afterimages.

**Results**

In experiment 1 we acquired decay curves for three types of recovery stimuli: a white blank screen, a single colored image and a barrage of colored images, all shown in a randomized order. Six subjects participated in the experiment and each gave a total of 195 judgments of afterimage pinkness (15 repetitions of each condition combination delay x type-of-recovery) (see methods).

The blank white screen did not have a noticeable effect on the intensity of afterimages in the first 800 milliseconds. The decay curves were shallow indicating no relevant reduction in the afterimage intensity (average drop 7%) (Figure 2A, blue). This was in a full agreement with past studies (Newsome, 1971; Virsu & Laurinen, 1977; Hansel & Mahmud, 1978).

However, an exposure to visual materials during recovery had a very different effect. We found that even a 200-millisecond exposure to a single colored image resulted in a detectable drop in afterimage intensity (33%), which was even more accentuated with longer exposures to recovery stimuli (46% at 400-milliseconds and 60% after 800 milliseconds exposure; Figure 2A, orange). A barrage of images showed a trend towards an even faster decay of after images (55% after 400 ms exposure) but ending with about the same reduction level as a single-colored image (59% at 800 milliseconds; Figure 2A, green). This indicates that afterimages are strongly washed-out by subsequent visual stimulation.

Importantly, the main finding of a fast decay of afterimages during subsequent visual stimulation was fully consistent across subjects (Figure 2B) (3x5 ANOVA repeated measures; main effects: F-values: 24.7 and 7.5; p-values: 0.000 and 0.000, respectively; interaction: F-value: 2.6; p-value: 0.016) and had huge effect sizes (at 800ms: d' color vs. blank = 6.38; d' barrage vs. blank = 4.67).

In experiment 2 we wanted to replicate the experiment 1 and expand the findings to gray-scale recovery stimuli. Six new subjects participated and were presented with the identical color and barrage conditions from experiment 1. In addition, a condition with gray-scale non-filtered images of natural scenes was presented as recovery stimuli (see methods).

The finding from experiment 1 was replicated (e.g., single full-colored image: 22% reduction after 200 ms exposure and 47% reduction after 800 ms exposure; Figure 2C). Also, gray-scale images washed out afterimages nearly as efficiently as colored images (37% reduction at 800 ms, lagging 10% behind the colored condition; Figures 2C, D).

The consistency among individual subjects was again high; 5/6 subjects clearly showed a fast decay of afterimages (Figure 2D orange). We confirmed the result that a barrage of colored images produced a slightly more effective cleanup than did a single colored image (32% and 55% reduction at 200 and 800 ms, exceeding by 10% and 8% the single colored condition, respectively; Figures 2C and E). The main effect of exposure time was again significant (ANOVA: F-value: 6.57 and 55.3; p-value: 0.0002), while others were not (all $p > 0.05$), indicating no statistically significant differences between the three types of images. And when computed against the blank screen from experiment 1, the effect sizes were again huge (at 800ms: d' color vs. blank = 3.05; d' barrage vs. blank = 3.18, d' gray vs. blank = 1.98).

Although experiment 2 replicated the main finding, the difference that barrage possibly made as opposed to a single image remained unclear; the difference replicated across experiments but did not appear significant. It was thus unclear whether barrage can accelerate cleanup or make it even overall more effective. Hence, to boost statistical power, we conducted experiment 3 with more subjects (n = 12) only presenting the two color and barrage conditions.

The results again replicated the main result from experiment 1; about one half of the afterimage was cleaned up quickly (F = 13.2; $p < 0.001$). Moreover, it became apparent that barrage could not produce higher levels of cleanup than single-colored images when presented for 800 ms (main effect: F = 2.2; p = 0.14; interaction: F = 0.76; p = 0.56) (Figure 2F, solid lines). This effect became marginally significant only when we pooled the data from all three experiments (n = 24) (main effect image: F = 2.95; p = 0.055) (Figure 2F, dotted lines). Therefore, the data suggested a *hard limit* for the degree to which afterimages could be cleared up by quick mechanisms within 800 ms: The limit lays at about one half of the initial afterimage intensity.

Finally, we found practice effects. At the beginning of the experiments, the subjects reported much less of the absolute cleanup level when compared to the middle of the experiments and the end. For example, at the 200-millisecond delay, twice as much cleanup was observed in the mid and late trials as compared to the early trails (see Figures 3A to C). This indicated that the visual system improves its afterimage cleanup skills.

## Discussion

Afterimages tend to last long. It is thus surprising to find that about 50% of the intensity of a green-induced afterimage can be erased rapidly—within less than a second. The prerequisite is that the subjects view a visual content rich with colors other than green. Thus, it appears that afterimages (partly) reflect corrections for the excess of green—a correction needed to achieve color constancy (Walsh 1995; Werner 2014). Therefore, afterimages arise when this corrective mechanism overcompensates by acting on the nothingness of a blank screen. Hence, cleanup of afterimages can be seen as a changed prior during Bayesian inference (Kass & Raftery 1995; Mathys et al. 2011; Foster 2011) about the true colors of an object. Such a change demands structured visual inputs and is hence ecological by its nature (Gibson 1979) and also embedded (Haugeland 1993). At the physiological level, we propose that those adjustments are underpinned by the fast adaptive mechanisms of neurons (Chung & Nelson 2002; Pearson 2000; Verhoef et al. 2008).

Our experiments also revealed gradual learning in the ability to clean up afterimages. Similarly, in other studies people needed time to learn adjustments to colors of sunglasses (Engel et al. 2016) and lenses (Tregillus et al. 2016). We propose that the fast adaptive mechanisms of neurons learn, and for that they rely on the mechanisms of intrinsic plasticity (Bliss & Lømo 1973; Andersen et al. 1977). Thereby, intrinsic plasticity is not assumed to alter the properties of neural networks directly but only indirectly by acting on the fast adaptive mechanisms first. It is only the fast adaptive mechanisms that adjust the neural network directly. This results in a tri-level adaptive hierarchy, as proposed by the theory of practopoiesis (Nikolić 2015; 2016; 2017)(Figure 3D). This theory obtains empirical support in the present findings of the fast cleanup of afterimages and the slow learning of the cleanup skill.

## Methods
**Participants**

A total of 24 participants took part in the study (9 male), ages 19-53 (mean 24). All had normal or corrected-to-normal vision and were not colorblind. All but one participant were naive to the purpose of the study, as surmised by a post-study questionnaire. All participants gave voluntary, informed consent to the study and were all reimbursed 10€ for their time.

**Stimuli**

The experiments were conducted in a dim, naturally-lit room. The stimuli were presented on an LCD monitor Samsung 2233RZ, suitable for psychophysical experimentation (Wang & Nikolić, 2011) and were programmed using Presentation 18.2. All stimuli were 1000 x 750 pixel photographs of real scenes, obtained with permission from public online sources. The adaptation stimuli were created from 120 unique colored photographs selected for their variety in color and visual complexity. Adobe Photoshop CS2 was used to impose a semitransparent green filter (RGB value 0,255,0 at 20% opacity) over each photograph. The stimuli were shown in randomized order.

The recovery stimuli included colored stimuli and black and white stimuli. The colored stimuli were eight color photographs selected however for their relative lack of color and complexity, as we wished to present unfiltered stimuli that visually engaged the color processing system, but which were not so saturated with color that they would inadvertently produce visual aftereffects, themselves. No colored filter was imposed. The black and white stimuli were eight black and white photographs selected for the predominance of white and light gray tones. The letter stimuli were presented entirely in grayscale.

The evaluation screen was a white 1000 x 750 pixel screen.

The response palette was presented as a horizontal array of nine distinct 75 x 75 pixel colored boxes spaced 45 pixels apart. The leftmost box was white (RGB values 255,255,255); with each successive box, from left to right, the green bit value decreased by five, such that the ninth and rightmost box was had RGB values 255,215,255. The left mouse button was used to select color that best matched the afterimage, and the trial ended immediately after the participant indicated a selection. All stimuli, including the color palette, were presented on a black (0,0,0) background.

The person hidden in the image in Figure 1A of the main text is Mona Lisa by Da Vinci.

**Responses**

Over the course of the experiment, 15 responses were collected for each unique experimental condition (195 responses in total). One response was excluded from a participant in experiment 2 because it was a significant outlier (more than 3 SD apart from the rest of responses) compared to the other responses in that category.

**Procedure**

The participants were told that a blank screen would appear at the end of each image series, that the screen may appear either white or a shade of light pink, and that a selection palette would appear on the monitor after the screen. They were instructed to use the mouse to select the shade of pink from the palette that best matched the color of pink they perceived the blank screen to be.

Each trial began with the presentation of 12 colored, green-filtered adaptation stimuli, each for 1000ms, for a total of 12s exposure to green-filtered images. In the 0 ms delay trials, the blank white 500ms evaluation screen immediately followed; in the trials with longer delays (200 to 800 ms in steps of 200 ms), the recovery stimuli were presented immediately after the adaptation stimuli, followed by the 500ms evaluation screen. Immediately after the evaluation screen, the selection palette was offered on the monitor. Participants then had five seconds to select a shade of pink from the palette that best matched the perceived color of the evaluation screen. After a selection was made, a blank black screen was presented for one second before initiating the next trial. The trials with different conditions were shown in a block-randomized order.

The experiment lasted a total of 57 minutes on average (min = 54.16, max = 59.21).

**Normalization**

Due to the between-individual variability of responses, we normalized the data on an individual basis prior to further data analysis. Each individual's score of pinkness was transformed relative to the level of pinkness observed at their 0-millisecond recovery delay (the baseline), whereby a score value equal to the baseline had a value of 1.

**Figure legends:**

**Figure 1. A)** An example afterimage hiding a person. The identity of the portrayed individual is revealed in the methods section. **B)** The time course of an experimental trial. The adaptation phase consists of images of natural scenes passed through 80% green filter. At each trial, one of the four different recovery stimuli were used for a variable duration. Finally, at the test stage subjects judged the amount of pinkness of the white screen, using the pinkness palette at the bottom of the screen.

**Figure 2.** Fast clean up of afterimages. **A)** Decay curves obtained in experiment 1: Changes in the intensity of afterimages along an 800 ms period while viewing either a blank white screen, non-filtered color images or a barrage of non-filtered colored images. Instead of showing the variations around means using only error bars, we show the full information about the variation by plotting in subsequent graphs the results of each individual subject. **B)** The results from A) shown for all six individual subjects. Color coding: the same as in A). **C)** Decay curves obtained in experiment 2: Changes in the intensity of afterimages while viewing either a gray scale image, non-filtered color images or a barrage of non-filtered colored images. **D)** The results from C) shown for all six individual subjects. Color coding: the same as in C). **E)** A scatter of afterimage intensities to directly compare unfiltered color images to barrages of images after 200 and 800 ms exposure; shown individually for all six subjects. Diagonal: equality line. **F)** Decay curves obtained in experiment 3: Solid lines: Changes in the intensity of afterimages while viewing either a non-filtered color images or a barrage for twelve subjects. Dotted lines: Color image and barrage data pooled across all three experiments.

**Figure 3.** Practice effects during the experiments. **A)** Solid blue line: Trial-resolved average changes of absolute after image judgments over the course of the experiment. Solid black line: Best fitted logarithmic function. **B)** The average for all three backgrounds that evoked fast adaptations during recovery shows that recovery from adaptation was both faster and overall more effective later in the experiment as compared to the early trials. **C)** The averages from B) decomposed for three different fast recovery conditions, shown for 200 and 800 milliseconds delays. Diagonal: equality line. **D)** Practopoiesis proposes a tri-level adaptive system: the slowest level (plasticity) trains the medium level (fast adaptive mechanisms), which in turn adjust the fastest part (neural networks). Presumably, the effects of both fast (cleanup) and slow (learning the cleaning skill) adaptive mechanisms have been at play in the present experiments.

Figure 1

A) 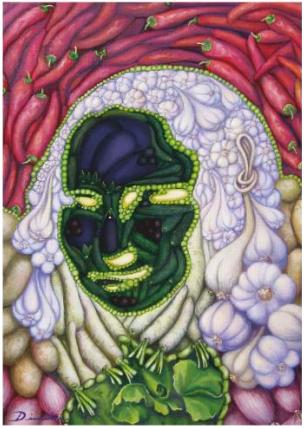

B) 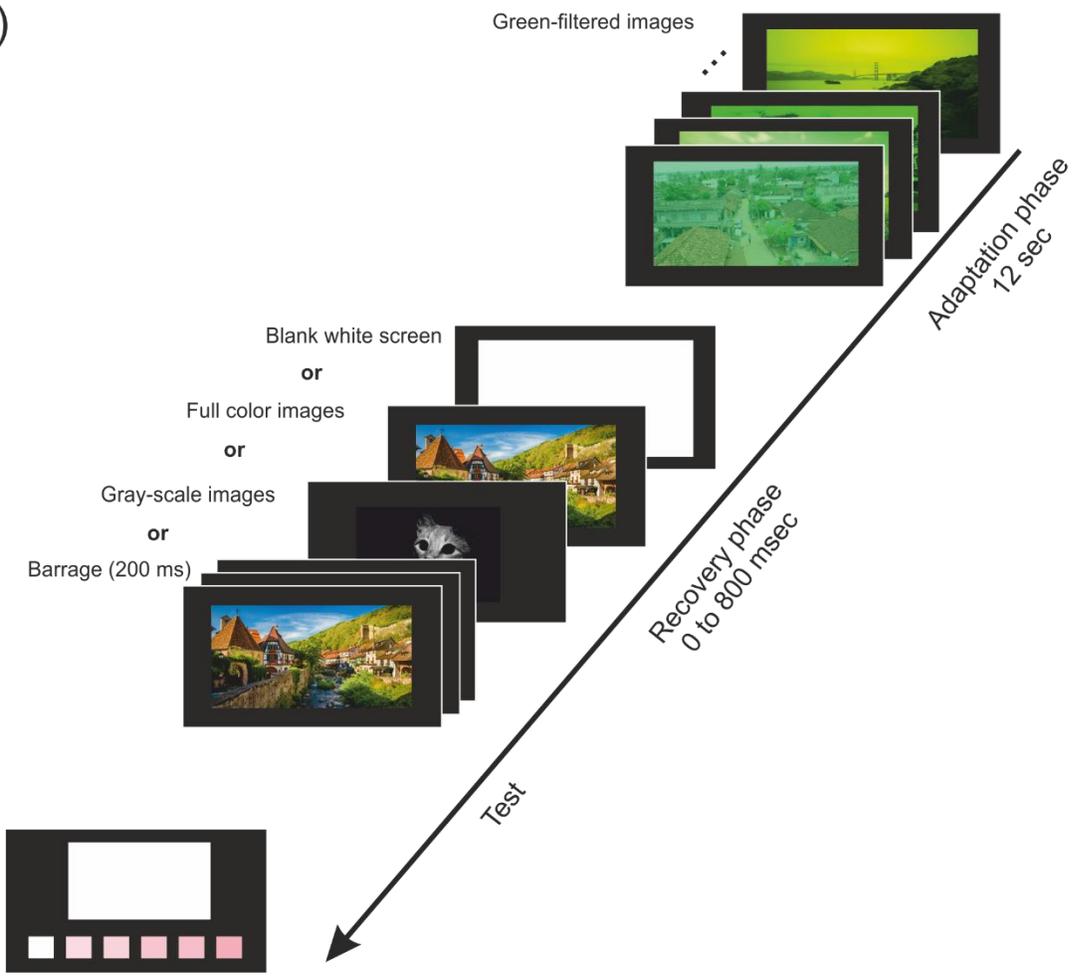

Figure 2

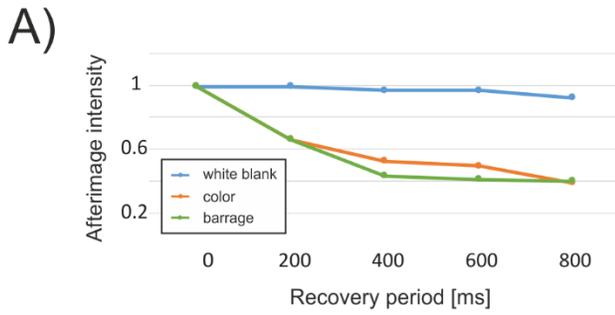
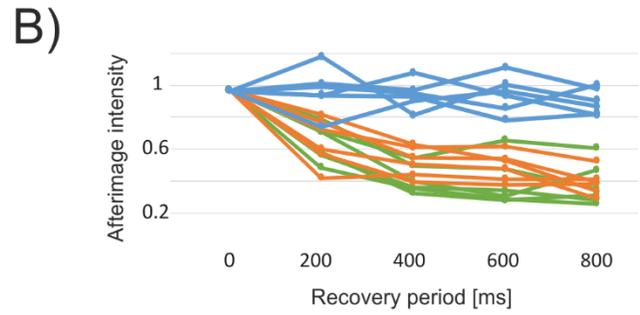
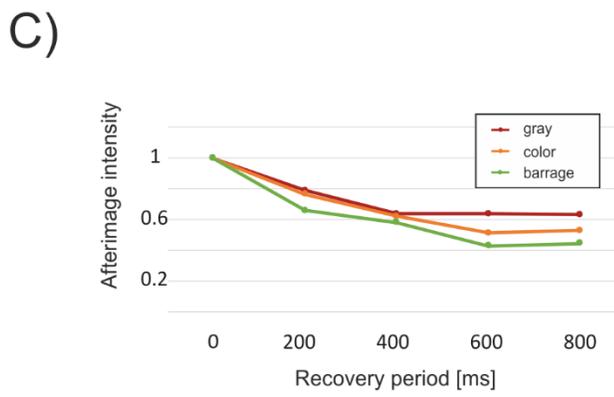
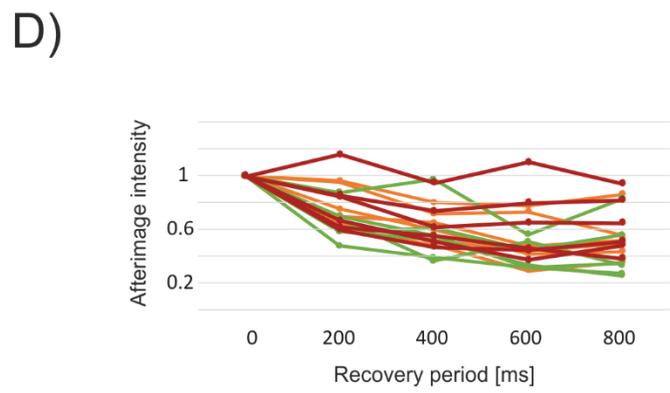
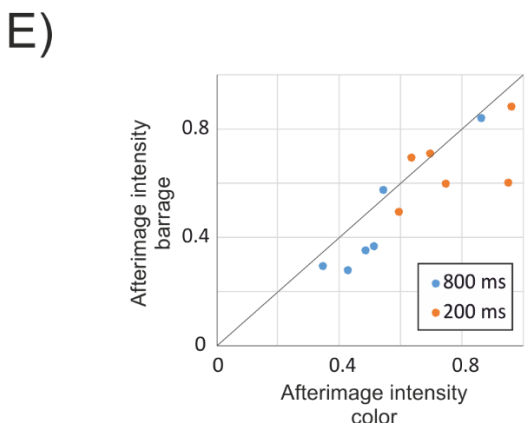
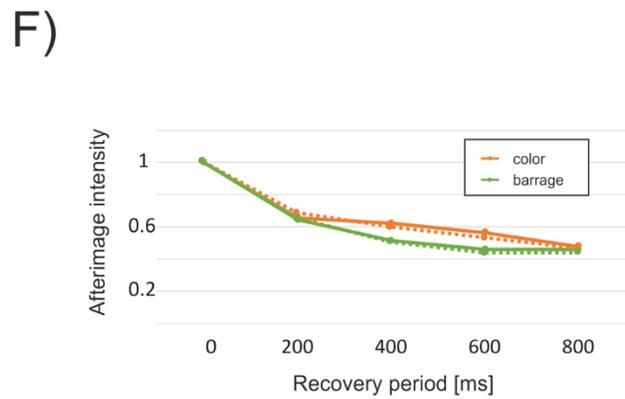

Figure 3

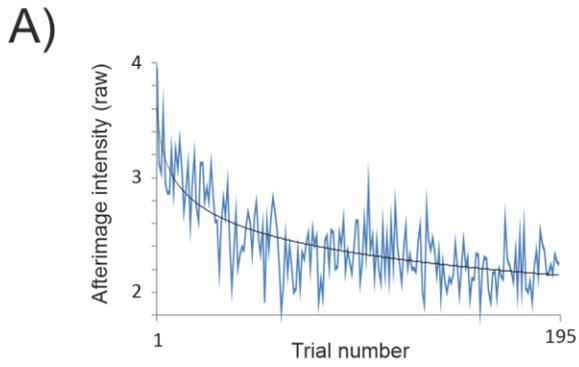
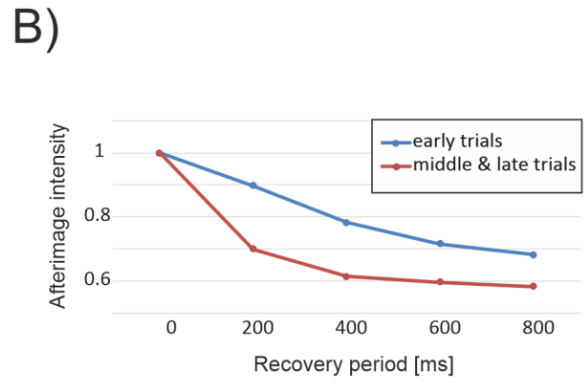
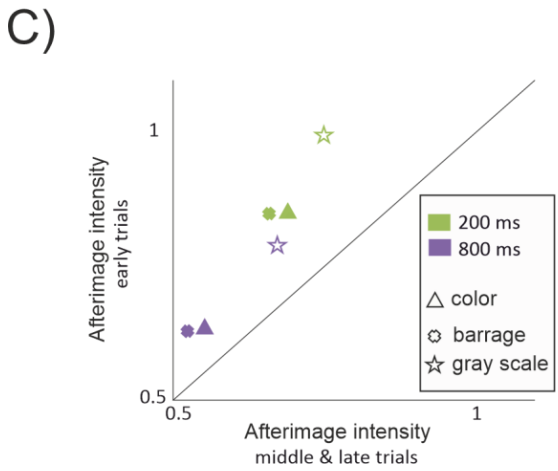
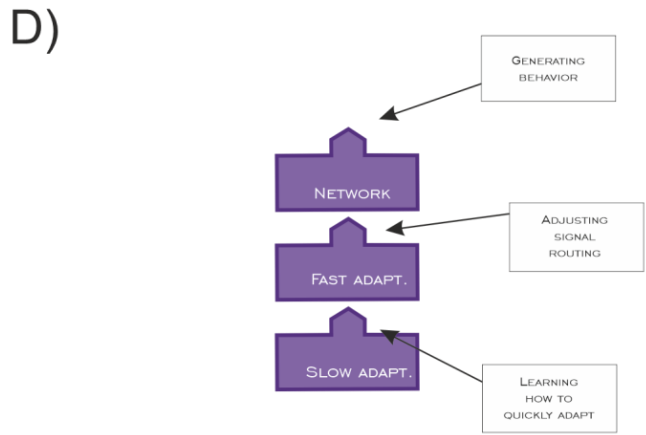